\documentclass[12pt]{JHEP}
\def\be{\begin{equation}}
\def\ee{\end{equation}}
\def\bea{\begin{eqnarray}}
\def\eea{\end{eqnarray}}
\title{The target space dependence of the Hagedorn temperature}

\author{Gianluca Grignani\\Dipartimento di Fisica and Sezione I.N.F.N., 
Universit\`a di Perugia, Via A. Pascoli I-06123, Perugia, Italia.
\email{E-mail:grignani$@$pg.infn.it}
\thanks{Work supported in part by INFN and MURST of Italy.}}

\author{Marta Orselli\\Dipartimento di Fisica and I.N.F.N. Gruppo
Collegato di Parma, Parco Area delle Scienze 7A I-43100 Parma, Italia.
\email{E-mail:orselli$@$fis.unipr.it}
\thanks{Work supported in part by MURST of Italy.}}

\author{Gordon W. Semenoff\\Department of Physics and Astronomy, 
University of British Columbia,
6224 Agricultural Road, Vancouver, British Columbia V6T 1Z1 Canada. 
\email{E-mail:semenoff$@$physics.ubc.ca}
\thanks{ Work supported in part by NSERC of Canada
}}
\abstract{

The effect of certain simple backgrounds
on the Hagedorn temperature in theories of closed strings is examined.  
The background of interest are a constant Neveu-Schwarz $B$-field,
a constant offset of the space-time metric and a compactified spatial 
dimension.
We find that the Hagedorn temperature of string theory depends on
the parameters of the background.  We comment on an interesting non-extensive
feature of the Hagedorn transition, including a subtlety with decoupling
of closed strings in the NCOS limit of open string theory and on the large
radius limit of discrete light-cone quantized closed strings.}


\preprint{UPRF-2001-19}

\begin{document}

One of the most interesting and general features of string theory is
its exponentially increasing density of states ~\cite{Hagedorn:1965st,
Huang:1970iq}.  If one considers an ensemble of weakly interacting
strings at finite temperature, this behavior of the density of states
is thought to lead either to a limiting temperature or a phase
transition.  The limiting temperature   is called the Hagedorn
temperature.  

In weakly coupled string theory this phenomenon can be understood in terms of
how the density of states in a multi-string system depends on  the energy.
Below a certain energy scale, the dominant contribution to the density of 
states in a system of many strings is a thermal distribution of multi-string 
states.
At a higher energy the statistically most
likely configuration changes from this thermal distribution to one which is 
dominated
by a single long string.  This leads to an interesting non-extensive behavior
of the thermodynamics at that point. 

Recently, there has been some interest in the Hagedorn transition in
background fields \cite{Ambjorn:2000yr}, 
particularly the behavior of open strings in the limit
which produces non-commutative open string (NCOS) theory 
\cite{Gubser:2000mf,Barbon:2001tm,Chan:2001gs}.  The phase
diagram of these systems has an exceedingly rich structure
\cite{Chan:2001gs}.  It also has interesting analogs in gauge theory
systems as was pointed out in a recent work \cite{Kogan:2001px}.

One advantage of the NCOS limit is that closed strings, and therefore
gravity, decouple \cite{Seiberg:2000ms} \cite{Gopakumar:2000na} from the open
string degrees of freedom.  This  avoids problems
which are expected to be inherent in trying to make a thermal ensemble in a
theory of quantum gravity.  Such a theory should suffer from
the Jeans instability at finite temperature - which simply means that hot flat
space is unstable, with the preferred state likely to be one where the energy 
density has collapsed into black holes whose Beckenstein-Hawking entropy is 
much 
greater than any
ordinary particle states.  It was argued in \cite{Atick:1988si} that
gravitational instability would make the Hagedorn transition of
ordinary string theories into a first order transition and that it should
actually occur at a temperature which is less than the Hagedorn temperature.
On the other hand, it was argued in \cite{Gubser:2000mf} that in the
NCOS theory, since gravity decouples, 
the transition is of second order and can be studied in the
context of weakly coupled string theory.

It was shown in \cite{Klebanov:2000pp}  that when the space is compactified, 
wrapped states of closed strings do not decouple in 
the NCOS limit.  These closed string states were used to construct wound
string theory in \cite{Danielsson:2000gi} and non-relativistic closed
strings \cite{Gomis:2000bd}.  One would expect that, in the limit as the 
compactification radius is large, the wrapped closed strings would couple 
more and more weakly and in the
infinite, de-compactified limit they would disappear from the spectrum.  
Indeed their energies do go to infinity.  However, we shall show in this 
paper that their Hagedorn temperature remains, that is, no matter how 
large that radius is, they still participate in the Hagedorn transition.  
This means that they should make a
contribution to the thermodynamic properties of the system.

We will begin by examining the effect of certain simple background
fields on the Hagedorn temperature in theories of closed strings.  The
NCOS limit is accessible within this family of backgrounds and can be
studied there.  This background is a space-time with one compact
dimension,
\begin{equation}
X^1\sim X^1+2\pi R
\label{identification}
\end{equation}
a Neveu-Schwarz $B$-field and space-time metric of the form
\begin{eqnarray}
G_{\mu\nu}= \left( \matrix{
-1+A^2 & -A & 0 & ... \cr -A & 1 & 0
& ... \cr 0 & 0 & 1 & ... \cr ....  & ...  & ...& ...\cr } \right)\ ,
~~~
B_{\mu\nu}= \left( \matrix{
0 &B  & 0 & ... \cr -B & 0 & 0
& ... \cr 0 & 0 & 0 & ... \cr ....  & ...  & ...& ...\cr } \right)\ ,
\label{metric}
\end{eqnarray}
where $A$ and $B$ are constants.  When $g_s=0$, closed string theory
is exactly solvable on this background.  Ordinarily, closed strings do not
couple to a constant $B$-field since, in the absence of D-branes it is gauge
equivalent to a constant electromagnetic field and closed strings do not carry 
electromagnetic charges.  Furthermore, they would not couple to $A$ since it
can be removed by a coordinate transformation.  However, when the
coordinate $X^1$ is compactified, neither the gauge transformation nor
the coordinate transformation are compatible with the identification
(\ref{identification}). 

When a spatial dimension is compactified, the 
wrapped modes of closed strings are indeed affected by $B$ which shifts their
energy by a constant.  
The shift of energy of the wrapped states can be understood by
considering a process where you make an wrapped closed string by
transporting the ends of an open string around the compact dimension
and then fusing them together.  Then, transporting the charged
endpoints of the open string in a constant $B$-field
involves precisely the energy shift which produces the chemical
potential for the resulting wrapped closed string state.

Similarly, on an un-compactified space, the parameter $A$ can
be shifted away by a re-definition of the coordinates, 
\bea
G_{\mu\nu}dX^\mu dX^\nu=
-dX^0dX^0+ (dX^1-AdX^0)(dX^1-AdX^0) + \ldots\nonumber
\eea
However, when $X^1$ is periodically identified, this re-definition is not a 
symmetry of the space-time.  In this case, the spectrum of closed strings also
couples to $A$ which shifts their energy by their momenta in the 1-direction.  For
example, a single bosonic closed string which wraps the compactified direction
$p$ times and which has $l$ quanta of momentum in that direction has energy spectrum 
\begin{equation}
P_0= -\frac{BRp}{\alpha'}- \frac{Al}{R}
+\sqrt{ \left(\frac{pR}{\alpha'}\right)^2+\left(\frac{l}{R}\right)^2+{\vec P}^{~2}+
\frac{2}{\alpha'}(N+\tilde N-2) }
\label{spectrum}
\end{equation}
(Here $N=\sum_{n=1}^\infty \alpha_{-n}\cdot\alpha_n$ and
$\tilde N=\sum_{n=1}^\infty {\tilde \alpha}_{-n}\cdot{\tilde
\alpha}_n$ with standard notation for oscillators following
\cite{Green:1987sp} and ${\vec P}^{~2}= P_iP^i$ with $i=2,...,D-1$.  
A similar formula for type II superstrings is
a straightforward generalization of (\ref{spectrum}).  It should be
supplemented by the appropriate level matching condition and, for the
fermionic string, the GSO projection.)

Our central result is that in the presence of $A$ and $B$ in the compact
space, the Hagedorn temperature is modified to be
\begin{equation}
T_H=T_H^0 \sqrt{ (1-A^2)(1-B^2) }
\label{hagedorn}
\end{equation}
where $T_H^0$ is the Hagedorn temperature of the string theory in the
limit where $A=B=0$.  For the bosonic string
$T_H^0=1/4\pi\sqrt{\alpha'}$ whereas it is $1/2\pi\sqrt{\alpha'}$ for the
type II superstring.

The formula (\ref{hagedorn}) has a remarkable feature.  As expected,
it depends on $A$ and $B$.  However, for fixed $A$ and $B$, it does
not depend on the compactification radius $R$.  This is surprising for
the following reason.  The main role of $A$ and $B$ in the string
spectrum is as chemical potentials for discrete momentum and wrapping
modes respectively, as can be seen for example from
the closed Bosonic string spectrum 
(\ref{spectrum}).

There is a region of the parameter space where $A$ and $B$ are between
0 and 1, away from their limiting values and where $R$ is very large
so that all wrapped states have a very large energy. In that case, at
temperatures just below $T_H$, practically no wrapped states are
excited in the thermal distribution.  However, since $T_H$ depends on
$B$, it must be wrapped states which condense at the Hagedorn
transition, in fact the resulting long string must wrap the compact
dimension.  An unwrapped long string could only become important at
the higher temperature
$T_H^0\sqrt{1-A^2}>T_H^0\sqrt{(1-A^2)(1-B^2)}$.  
Thus we see that, in the limit where $R$ is very large, when the
temperature $T_H$ is reached, there is a catastrophic process where
dominant configurations in the ensemble go from a thermal distribution
of multi-string states with zero wrapping to a single long string
which wraps the compact dimension.

In a thermal ensemble where the total energy is proportional to the
volume, there is certainly sufficient energy to produce such a long
string whose energy only scales like its length.  Then, the
$R$-dependence of the total energy, which grows linearly in $R$ if the
temperature is held fixed as $R$ is changed, is similar to the energy
dependence of a wrapped string which also scales linearly with $R$.

In \cite{Klebanov:2000pp} it was noted
that, when the compactified dimension has finite radius,  
the wrapped closed string states do not
decouple in the NCOS limit.   These wrapped states should get infinitely large
energy in the limit where the radius of the compact dimension is taken to 
infinity.  However, we see that, no matter how large that radius is, the
closed strings still participate in the Hagedorn transition.   The phase
transition of open strings in the decompactified
NCOS limit is thought to be of second order
\cite{Gubser:2000mf}. We see that, if the radius is very large but
finite, the closed string Hagedorn behavior makes it a first order transition.

It is clear from (\ref{hagedorn}) that there are limiting values of
both background fields $A$ and $B$.  The critical value of $B$ is
where the NCOS limit is found.  A whole family of NCOS limits 
should arise in our model by changing $A$ within its limiting values. 
As can be easily seen, a T-duality transformation along the compactified 
direction~\cite{Buscher:1988qj,Buscher:1987sk}, 
simply interchanges the role of $B$ and $A$, 
$B\leftrightarrow A$. Since the Hagedorn temperature
(\ref{hagedorn}) is symmetric under this interchange, it is self-dual.

The limiting
value of $A$ is similarly interpreted as the DLCQ limit of the closed
string theory.  In fact, it can be seen explicitly that taking $A=1$
in (\ref{spectrum}) (with the appropriate rescaling of $R$) reproduces
the discrete light-cone quantization (DLCQ) 
spectrum of closed strings in a $B$-field that was discussed
in \cite{Grignani:2001hb}.  There, it had an interesting
interpretation in terms of covers of a torus that are expected to be
found in the weak coupling limit of the matrix model of M-theory
\cite{Grignani:1999sp,Grignani:2000zm}.
The result of the present paper implies a curious non-decoupling in the 
DLCQ limit of closed strings.  This is another limit of string theory 
which is described by a gauge
theory, the matrix model, which does not involve gravity.  We found
in \cite{Grignani:2001hb} that the $B$-field couples to the thermodynamic
partition function of both free type II superstring theory and the 
matrix string.  Indeed, the Hagedorn temperature there is also modified
by a factor of $\sqrt{1-B^2}$, with no reference to the light-cone radius $R$.
This poses a subtlety for discrete light-cone quantization of strings.

The energy spectrum in (\ref{spectrum}) is straightforward to obtain
from canonical quantization of the string.  The nature of the high
energy density of states with such a spectrum was discussed in detail
 \cite{Deo:1989bv}. 
In fact there are several ways of finding the Hagedorn temperature.
One is to estimate the asymptotic density of states $\rho(E)\sim
\exp(\beta_H E)$ and find the coefficient in the exponential
$\beta_H=1/k_B T_H$ where $k_B$ is the Boltzmann constant.  In this
paper we are using units where $k_B=1$.  Another \cite{Atick:1988si}
is to examine the spectrum of the string theory and see where a new
tachyonic state appears.  In all known cases, this temperature
coincides with the Hagedorn temperature.  Finally, the Hagedorn
temperature can be defined as that temperature where, in Euclidean
space, the vertex operator
$$
e^{2\pi i T X^0}
$$
becomes a relevant operator.

It is this last criterion where the Hagedorn transition is seen to be
analogous to the Berezinsky-Kosterlitz-Thouless (BKT) transition in
the 2-dimensional X-Y-model, a parallel which has been drawn many
times in the literature \cite{Sathiapalan:1987db} \cite{Kogan:1987jd}.
In fact  our present model could have an interesting analog in
coupled X-Y-models where the metric and $B$-field couple the two angular
degrees of freedom,
$$
S=-\frac{1}{4\pi\alpha'}\int d^2\sigma  \partial_a X^\mu\left(
G_{\mu\nu}\delta^{ab}-B_{\mu\nu}\epsilon^{ab}\right)\partial_b X^\nu
$$
The BKT transition involves the condensation of vortices.  It is easy
to see that the transition temperature is modified by $A$ and $B$ in
the same way as the Hagedorn temperature\footnote{Note that the
temperature at which the BKT transition occurs is not universal.
Here, by BKT temperature, we mean the temperature at the zero coupling
limit of the line of critical points.}.  The analog of the catastrophic
behavior which we discussed at the Hagedorn temperature is a
condensation of vortices of one of the variables $X^0$ induced by the
$B$-coupling to $X^1$ in a state where the density of these vortices was
zero just before the transition.  It is possible that this process
could be experimentally visible in Josephson junction
arrays\footnote{We thank Professor P. Sodano for discussions on this
point.}.

\noindent
{\bf Derivation of $T_H$:}

The free energy of a gas of relativistic Bose particles is
\begin{equation}
F=\frac{1}{\beta}{\rm Tr}\ln\left( 1-e^{-\beta P_0}\right)=
-\sum_{n=1}^\infty
\frac{1}{n\beta}{\rm Tr} e^{-n \beta P_0} 
\label{particle}
\end{equation}
Equation (\ref{particle}), can be used to derive the bosonic string
free energy at one loop, by the standard procedure of computing
the sum of free energies of the particles in the string spectrum.
Canonical quantization of the string in the light-cone gauge give
the energy spectrum (\ref{spectrum}) together with the level
matching condition $\tilde N -N= pl$.  
To obtain the free energy of the bosonic string we use the integral
identity 
$$
\int_0^\infty dt e^{-x t^2-y/{t^2}}=
\frac{1}{2}\sqrt{\frac{\pi}{x}}e^{-2\sqrt{xy}}
$$
where 
$$
t^2=1/\tau_2,~~~~~ x=\frac{n^2\beta^2}{4\pi \alpha'},
~~~~
y=\pi \alpha' \left(\frac{R^2p^2}{{\alpha'}^2}+\frac{l^2}{R^2}+
{\vec P}^2 \right)
$$

We also enforce the level matching condition with a Lagrange 
multiplier $\tau_1$ to obtain the free energy of the bosonic string
\begin{eqnarray}
&&F=-\sum_{n,p,l}\int_0^{\infty} \frac{d\tau_2}{2\tau_2}\int_{-1/2}^{1/2}
\frac{d\tau_1}{(4\pi^2\alpha'\tau_2)^{13}}
\left(\frac{\alpha'\tau_2}{R^2}\right)^{1/2}
\left|\eta(\tau)\right|^{-48}\cr
&&\exp\left[-\frac{\beta^2 n^2}{4\pi\alpha'\tau_2}-
\pi\alpha'\tau_2\left(\frac{l^2}{R^2}+\frac{R^2 p^2}{\alpha'^2}\right)
-2\pi i\tau_1 p l+ n\beta {B}\frac{Rp}{\alpha'}
+n\beta {A}\frac{l}{R}\right]
\label{free}
\end{eqnarray}
The temperature independent $n=0$ term gives the vacuum energy, 
$i.e.$ the cosmological constant contribution,
the other terms give the relevant thermodynamic potential.

To perform the integration over $\tau_1$ it is useful to 
rewrite the Dedekind eta function in terms of a series
as in \cite{Green:1987sp}. One has
\begin{equation}
\left|\eta(\tau)\right|^{-48}=e^{4\pi\tau_2}\left|\prod_{m=1}^\infty
\left(1-e^{2\pi i \tau m}\right)\right|^{-48}
\label{etad}
\end{equation}
and  
\begin{equation}
\prod_{d=1}^\infty
\left(1-z^d\right)^{-24}\equiv \sum_{r=0}^\infty d(r) z^r
\label{prod}
\end{equation}
where $z=\exp(2\pi i \tau)$.

Using (\ref{etad}) and (\ref{prod}),
the $\tau_1$ integral in (\ref{free}) can be easily performed
\begin{equation}
\sum_{r,r'=0}^\infty d(r) d(r')e^{-2\pi\tau_2(r+r')}\int_{-1/2}^{1/2}
{d\tau_1}e^{2\pi i\tau_1 (r-r'+p l)}=
\sum_{r=0}^\infty d(r) d(r+pl)e^{-2\pi\tau_2(2r+pl)}
\label{sum}
\end{equation}
The coefficient $d(r)$ is given by 
\begin{equation}
d(r)=\frac{1}{2\pi i}\oint \frac{G(z)}{z^{r+1}}
\label{density}
\end{equation}
where 
\begin{equation}
G(z)\equiv\sum_{r=0}^\infty d(r) z^r={\rm Tr} z^N=\prod_{r=1}^\infty
\left(1-z^r\right)^{-24}
\end{equation} 
The generating 
function $G(z)$ vanishes rapidly for $z\to 1$, while if $r$ is very large
$z^{r+1}$ is very small for $z<1$. Consequently, for large $r$
there is a sharply defined saddle point for $z$ near 1.
Following \cite{Huang:1970iq} one gets
\begin{equation}
d(r)\sim r^{-27/4}e^{4\pi\sqrt{r}}
\label{asympt}
\end{equation}
In the $\tau_2\to 0$ limit the sums are dominated by those integers
for which $r,\ l$ and $p$ are such that $r$ and $r+pl$ are big, 
so that (\ref{asympt}) 
could be used for $d(r+pl)$. Moreover, the dominant term is obtained by
setting $n=1$.
Then for $\tau_2\sim 0$
we could use a saddle point procedure for the variables
$r$, $l$ and $p$ to evaluate the sums
\begin{equation}
\sum_{l,p}\sum_{r=0}^{\infty}  r^{-27/4} (r+pl)^{-27/4}
e^{4\pi\left(\sqrt{r}+\sqrt{r+pl}\right)-2\pi\tau_2(2r+pl)-
\pi\alpha'\tau_2\left(\frac{l^2}{R^2}+\frac{R^2 p^2}{\alpha'^2}\right)
+\beta {B}\frac{Rp}{\alpha'}
+\beta {A}\frac{l}{R}}
\label{saddle}
\end{equation}
The equations that are to be solved to find the maximum of the exponent
are 
\begin{eqnarray}
& &\frac{1}{\sqrt{r}}+\frac{1}{\sqrt{r+pl}}-2\tau_2 =0 \cr
& &\beta \frac{{A}}{R} -2\pi \alpha'\tau_2 \frac{l}{R^2}+
2\pi \frac{p}{\sqrt{r+pl}} -2\pi \tau_2 p =0 \cr
& &\beta \frac{{B}R}{\alpha'} -2\pi \alpha'\tau_2 \frac{pR^2}{\alpha'^2}+
2\pi \frac{l}{\sqrt{r+pl}} -2\pi \tau_2 l =0
\label{max}
\end{eqnarray}
The solutions for $p, l$ and $r$ read
\begin{eqnarray}
& &p=\frac{\tau_2 r}{1-2\tau_2 \sqrt{r}}\frac{\beta }{2\pi R}
\left(\frac{{A}}{\tau_2 \sqrt{r}}-{A}-{B}\right)\cr
& &l=\frac{\tau_2 r}{1-2\tau_2 \sqrt{r}}\frac{\beta R}{2\pi \alpha'}
\left(\frac{{B}}{\tau_2 \sqrt{r}}-{B}-{A}\right)\cr
& &\sqrt{r}=\frac{1}{2\tau_2}\left(1\pm \sqrt{\frac{1+
\frac{\beta ^2}{16 \pi ^2 \alpha'}({A} -{B})^2}
{1+\frac{\beta ^2}{16 \pi ^2 \alpha'}({A} +{B})^2}}\right)
\label{sol}
\end{eqnarray}
To obtain the well-known solution for ${A}={B}=0$,
i.e. $\sqrt{r}=1/\tau_2 $, we must choose the $+$ sign 
in the last equation.

Substituting the solutions (\ref{sol}) in (\ref{free})
the exponent becomes
\begin{equation}
\frac{2\pi}{\tau_2}\left\{1-\frac{\beta ^2}{8 \pi ^2 \alpha'}
-\frac{\beta ^2({A}^2 +{B}^2)}{16 \pi ^2 \alpha'}+
\sqrt{\left(1-\frac{\beta ^2({A} +{B})^2}{16 \pi ^2 \alpha'}\right)
\left(1-\frac{\beta ^2({A} -{B})^2}{16 \pi ^2 \alpha'}\right)}\right\}
\end{equation}
It is not difficult to see that this exponent vanishes
when $T=T_H=1/\beta_H$, where
\be
T_H=\frac{\sqrt{(1- A^2)(1- B^2)}}{4\pi\sqrt{\alpha'}}
\label{thab}
\ee
The Hagedorn temperature does not depend on the 
compactification radius and is smaller then the Hagedorn
temperature in the absence of $A$ and $B$.
It is interesting to notice that $A$ and $B$ play the role of 
the chemical potentials for the quantized momenta and
winding modes in the compactified direction, respectively. 
In fact, the formula for the chemical potential 
dependent Hagedorn temperature derived in
\cite{Deo:1989bv} can be shown to be
identical to (\ref{thab}). In  
\cite{Deo:1989bv} the reduced chemical potentials for
the quantized momenta and
winding modes 
$\bar\mu=\beta\mu$ and $\bar\nu=\beta\nu$ were used.
Thus, performing the necessary
rescaling by $\beta$, one arrives,
at the Hagedorn temperature 
\begin{equation}
T_H=\frac{1}{4\pi\sqrt{\alpha'}}\sqrt{(1-(\nu R)^2)
\left(1-\left(\frac{\mu\alpha'}
{R}\right)^2\right)}
\end{equation}
Comparing this with(\ref{thab}) we can identify 
$$
\nu\equiv
\frac{A}{R}~,~~~~~~~~\mu\equiv \frac{R B}{\alpha'}
$$
In terms of $A$ and $B$ any dependence on the compactification radius
$R$ disappears.
This independence on $R$
is remarkable since it must hold even if the compactification radius is
arbitrarily large.  Of course, without the compactification in the
first place, $T_H$ would be independent of $A$ and $B$ and would be the usual
closed string value $1/4\pi\sqrt{\alpha'}$.  This non-commutativity of
compactifying and going to the Hagedorn temperature is a result of the
exponential growth of the density of states of the string which is
independent of compactification radius. At the Hagedorn temperature, the 
thermal distribution of string states is unstable and the most favorable
configuration is one long string that contains all of the energy.  In order
to know about the $A$ and $B$ fields, 
this long string must wrap the compactified
light-like direction.  Because of this non-extensive behavior, it always
has enough energy to do that, no matter how large the radius $R$.

Similar expressions with similar conclusions can be reached for the case
of the type II superstring and the result is given in (\ref{hagedorn}).

\end{document}